\documentclass{iopart}

\usepackage{iopams}

\begin{document}

\title[Relativistic and QED corrections to the $2p\sigma_u(v=1)$
state of the $\mbox{H}_2^+$]{Relativistic and QED corrections to the
$2p\sigma_u(v=1)$ vibrational state of the $\mbox{H}_2^+$ molecular ion}

\author{J.~Carbonell}
\address{
Laboratoire de Physique Subatomique et de Cosmologie\\
53, avenue des Martyrs, 38026 Grenoble, France}
\author{R.~Lazauskas}
\address{
CEA/DAM/DPTA Service de Physique Nucl\'eaire\\
B.P.~12, F-91680, Bruy\`eres le Ch\^atel, France}
\author{V.I.~Korobov}
\address{Joint Institute for Nuclear Research\\
141980, Dubna, Russia}

\begin{abstract}
Relativistic and QED corrections to the recently discovered first
vibrational $2p\sigma_u$ state are presented. This state has an extremely
small nonrelativistic binding energy $E_B=1.085045252(1)\times10^{-9}$
a.u. Its wave functions has a maximum at $R\approx100$ a.u. and extends up
to several hundreds. It is shown that this state does not disappear if
higher order relativistic and QED corrections, including the
Casimir--Polder effect, are taken into account.
\end{abstract}


\section{Introduction}

We have recently reported \cite{Car02,Car03,Laz02} on the possible
existence of a new bound state of the $\mbox{H}_2^+$ molecular ion. It
corresponds to  the first excitation of the $2p\sigma_u$ vibrational
state, i.e.\ a state  with total three-body orbital angular momentum $L=0$
and with proton-proton spin coupled to $S_{n}=1$. Its binding energy is
extremely small, $E_B=1.085045252(1)\times10^{-9}$ a.u. (atomic units)
with respect to the $p$--H dissociation threshold, and its wave function
extends over several hundreds a.u. Although not being populated, it
results into a huge $p$--H scattering length value of $a_t\approx750$ a.u.,
which dominates the low energy $p$--H scattering. Apart from its exotic
character, the very existence of such a state can dramatically modify the
$p+\mbox{H}\to \mbox{H}_2^++h\nu$ reaction rate and consequently can
explain the $\mbox{H}_2$ molecular abundance
\cite{Smukstaras,Smukstaras1}. Our previous calculations were done in the
framework of nonrelativistic quantum mechanics. Given the smallness of
its binding energy, it is necessary to undertake a systematic study of the
relativistic corrections in order to see to what extent they modify, or
even question, the results previously found. This is the aim of the
present paper.

At a first glance it may be expected that at very large distances a change
of the asymptotic potential due to the Casimir--Polder (or retardation)
effect can lead to disappearance of this weakly bound state. For the
charge--dipole case the asymptotic potential at distances $r\gg 100$ a.u.\
has a form \cite{Ber76}
\begin{equation}\label{ret}
V(R) = -\frac{\alpha_d}{2R^4}
   \left(
      1-\frac{11\alpha}{2\pi}\frac{m_e}{m_p}\frac{1}{R}
   \right)
\end{equation}
where $\alpha_d$ is the electric dipole polarizability of an atom (for the
hydrogen atom $\alpha_d=9/2$) and $\alpha\approx 1/137$ is the fine
structure constant. The first term is the classical polarization
potential, which results from the instant Coulomb interaction and is
already included in the nonrelativistic solution. The second term
represents the two transverse photon exchange between a neutral system
(hydrogen atom) and a distant charged particle (proton). In order to
evaluate the influence of this contribution on the binding energy one may
use the value of the retardation potential --- second term in
Eq.~(\ref{ret}) ---  at the maximum of the bound state wave function,
$R_{max}\approx100$ a.u., that gives:
\[
V_{\rm ret}(r_{max})=
    -(9/4)(11\alpha/2\pi)(m_e/m_p)R_{max}^{-5}
    \approx10^{-15}\hbox{ a.u.}.
\]
Thus the change of the $2p\sigma_u(v=1)$ binding energy is of the order
$\Delta E_{\rm ret}=E_B\times10^{-6}$. The smallness of this quantity
suggests that relativistic and QED corrections can be treated using the
standard perturbation technique for bound states, and eventually at some
stage should take in the Casimir--Polder effect.

\section{Variational calculation of the nonrelativistic solution}

The numerical calculations of the bound state wave function
have been performed using the variational
approach described in details in \cite{Kor00}. The variational
wave function for an $S$ state has the form,
\begin{equation}\label{expansion}
\begin{array}{@{}l}
\displaystyle
\Psi(\mathbf{r}_1,\mathbf{r}_2) = \sum_{i=1}^{\infty}
    \Big\{
       U_i\,{\rm Re}\bigl[e^{-\alpha_i r_1-\beta_i r_2-\gamma_i r}\bigr]
\\[4mm]\displaystyle\hspace{25mm}
      +W_i\,{\rm Im}\bigl[e^{-\alpha_i r_1-\beta_i r_2-\gamma_i r}\bigr]
    \Big\}-(1\leftrightarrow2),
\end{array}
\end{equation}
where $\mathbf{r}_1$ and $\mathbf{r}_2$ are the position vectors of the
electron with respect to two protons. Complex parameters $\alpha_i$,
$\beta_i$ and $\gamma_i$ are generated in a quasi-random manner:
\[
\begin{array}{@{}l}
\displaystyle
\alpha_i =
   \left[\left\lfloor
            \frac{1}{2}i(i+1)\sqrt{p_\alpha}
         \right\rfloor(A_2-A_1)+A_1\right]
\\[4mm]\displaystyle\hspace{15mm}
 +i\left[\left\lfloor\frac{1}{2}i(i+1)\sqrt{q_\alpha}
         \right\rfloor(A'_2-A'_1)+A'_1\right],
\end{array}
\]
$\lfloor x\rfloor$ designates the fractional part of $x$, $p_\alpha$ and
$q_\alpha$ are some prime numbers, $[A_1,A_2]$ and $[A'_1,A'_2]$ are real
variational intervals which need to be optimized. Parameters $\beta_i$ and
$\gamma_i$ are obtained in a similar way.

In order to get the accurate result we use 5 sets of the basis functions
of the type (\ref{expansion}), for which the variational parameters have been
searched independently. The proton-to-electron mass ratio,
$m_p=1836.152701\,m_e$, has been adopted for this calculations.

\begin{table}[t]
\begin{center}
\begin{tabular}{c@{\hspace{12mm}}l}
\hline\hline
Number of state ($N$) & \hspace*{-4mm}Binding energy $E_B$ (in a.u.) \\ \hline
2000 & $1.08504520\times10^{-9}$ \\
2500 & $1.085045237\times10^{-9}$ \\
3000 & $1.0850452464\times10^{-9}$ \\
3500 & $1.0850452494\times10^{-9}$ \\ \hline
$\infty$ & $1.085045252(1)\times10^{-9}$ \\
\hline\hline
\end{tabular}
\end{center}
\caption{Convergence of the binding energy (in a.u.) for the
$2p\sigma_u(v=1)$ state with respect to a number of basis functions}\label{Table1}
\end{table}

In Table \ref{Table1} we present the convergence of the computed binding
energies as a function of $N$, number of the basis functions. One can see
that the nonrelativistic binding energy for this weakly bound state has a
relative accuracy of $10^{-9}$, what is compliant with the requirements of
the precise spectroscopy. The next question is how to improve this value
by taking into account the corrections imposed by a relativistic theory
and QED. These aspects will be discussed in the following sections. As it
was demonstrated in the introduction, these corrections can be evaluated
using the standard perturbation expansion over the parameter
$\alpha\approx 1/137$, which can be derived from the nonrelativistic QED
effective field theory \cite{Pac98}.

\section{Corrections due to the Breit--Pauli Hamiltonian}

The Breit--Pauli Hamiltonian provide us with the relative $\alpha^2$ order
corrections with respect to the nonrelativistic energy of a state.

The major contribution comes from the relativistic correction for the
bound electron,
\begin{equation}\label{rel_electron}
\delta E_{rc}^{(2)} = \alpha^2\!\left\langle\!
         -\frac{\mathbf{p}_e^4}{8m_e^3}
         +\frac{4\pi}{8m_e^2}
           \left[\delta(\mathbf{r}_1)
                +\delta(\mathbf{r}_2)
           \right]
      \right\rangle.
\end{equation}
The other term of the Breit--Pauli Hamiltonian which has to be considered,
is the transverse photon exchange contribution, which reads:

\begin{equation}\label{trans}
\begin{array}{@{}l}
\displaystyle
\delta E_{tr\mbox{-}ph}^{(2)} = \frac{\alpha^2}{2M_p}
  \left\langle
     \frac{\mathbf{p}_e\mathbf{p}_1}{r_1}
    +\frac{\mathbf{r}_1(\mathbf{r}_1\mathbf{p}_e)\mathbf{p}_1}{r_1^3}
    +(1\leftrightarrow2)
  \right\rangle
\\[4mm]\displaystyle\hspace{15mm}
-\frac{\alpha^2}{2M_p^2}
  \left\langle
     \frac{\mathbf{p}_1\mathbf{p}_2}{R}
    +\frac{\mathbf{R}(\mathbf{R}\mathbf{p}_1)\mathbf{p}_2}{R^3}
  \right\rangle.
\end{array}
\end{equation}
The remaining recoil corrections are negligibly small compared to
uncertainty in the relativistic correction for the bound electron.

Beyond these terms, we have included as well the correction due to the
finite size of the proton,
\begin{equation}\label{fsc}
\delta E_{\rm nuc} = \frac{2\pi(R_p/a_0)^2}{3}
\Bigl\langle \delta(\mathbf{r}_1)+\delta(\mathbf{r}_2) \Bigr\rangle,
\end{equation}
where $R_p=0.862(12)\mbox{ fm}$ is the  proton root-mean-square radius.

\begin{table}[t]
\begin{center}
\begin{tabular}{lr@{}l}
\hline\hline
$\Delta E_{nr}$           & $-$1.&085$\>045\>252\times10^{-9}$ \\
\hline
$\Delta E_{rc}$           &    0.&003$\>285\>2(4)\times10^{-9}$ \\
$\Delta E_{tr\mbox{-}ph}$ &    0.&000$\>013\>371\times10^{-9}$ \\
$\Delta E_{nuc}$          & $-$0.&000$\>000\>067\times10^{-9}$ \\
\hline
$\Delta E_{\alpha^2}$     &    0.&003$\>298\>5(4)\times10^{-9}$ \\
\hline\hline
\end{tabular}
\end{center}
\caption{The Breit--Pauli Hamiltonian corrections (in a.u.) to the binding
energy of the $2p\sigma_u(v=1)$ state}\label{Table2}
\end{table}

As can be seen in the Table \ref{Table2} the relativistic correction to
the binding energy is of the order $\sim10^{-3}$, what is in agreement
with the work of Howells and Kennedy \cite{How90}. These authors studied
the relativistic corrections for the high vibrational states of the
$1s\sigma_g$ series in H$_{2}^{+}$. They found that the Breit--Pauli
relative contribution to the binding energy of the weakly bound states is
of the order $10^{-3}$, while in the case of low vibrational states it
constitutes only a $\sim10^{-5}$ part of the binding energy. The value of
$\Delta E_{nr}$ in Table \ref{Table2} agrees well with our previous
estimate \cite{Laz02}, based on the simplified approach of \cite{How90}.

The uncertainty in the relativistic correction for the bound electron is
considerably larger than other uncertainties in $\alpha^2$ corrections.
This is due to the strong cancelation between the correction terms for the
$\mbox{H}_2^+$ molecular ion and the ground state of the hydrogen atom.
The very accurate variational solution, providing the accuracy for the
nonrelativistic energy to be $\sim10^{-18}$ a.u., is still not enough to
get the precise value for this relativistic contribution.

\section{Radiative and higher order relativistic corrections}

The complete spin-independent contribution of order $\alpha^3$ and
$\alpha^3(m/M)$ has the form \cite{Pac98,Yel01},
\begin{equation}\label{alpha3}
\begin{array}{@{}l}
\displaystyle
\delta E^{(3)} = \alpha^3\sum_{i=1,2} \biggl[
\frac{4}{3}
    \left(
       \!-\!\ln\alpha^2\!-\!\beta(L,v)
       \!+\!\frac{5}{6}-\frac{1}{5}
    \right)
    \langle\delta(\mathbf{r}_i)\rangle\\[4mm]
\displaystyle\hspace{14mm}
   +\frac{2}{3M_p}\left(\!-\!\ln\alpha\!-\!4\,\beta(L,v)
                                          \!+\!\frac{31}{3}\right)
        \langle\delta(\mathbf{r}_i)\rangle
        -\frac{14}{3M_p}Q(r_i)\biggl],
\end{array}
\end{equation}
where
\begin{equation}\label{Bethe_log}
\beta(L,v) =
   \frac{
   \left\langle
      \mathbf{p}_e(H_0\!-\!E_0)\ln\left((H_0\!-\!E_0)/R_\infty\right)
      \mathbf{p}_e
   \right\rangle}
   {4\pi\left\langle
      \delta(\mathbf{r}_1)+\delta(\mathbf{r}_2)
   \right\rangle}
\end{equation}
is the Bethe logarithm, $H_0$ is the three--body nonrelativistic
Hamiltonian, $\mathbf{p}_e$ is the electron momentum operator and $Q(r)$
is the $Q$-term introduced by Araki and Sucher \cite{as},
\[
Q(r) = \lim_{\rho \to 0} \left\langle
            \frac{\Theta(r - \rho)}{ 4\pi r^3 }
      + (\ln \rho + \gamma_E)\delta(\mathbf{r}) \right\rangle.
\]

In calculating the $2p\sigma_u(v=1)$ state of $H_2^+$, the Bethe logarithm
was taken equal to the hydrogenic limit, namely,
$\beta(^3\!S,1)\approx2.9841$. This is justified since the electronic wave
function for the $2p\sigma_u(v=1)$ state to a good extent can be
approximated by the antisymmetrized hydrogenic wave function:
$\psi_e(r_1,r_2;R)=(1/\sqrt{2})(\psi_{\rm H}(\mathbf{r}_1)-\psi_{\rm
H}(\mathbf{r}_2)))$. On the other hand, this accuracy is sufficient to get
a relevant estimate of the $\alpha^3$ order radiative correction.

Our calculations  include also the $\alpha^4$ order corrections
in a non-recoil limit. Among them are the one-loop self-energy and vacuum
polarization corrections for the bound electron (next to the leading term
in $\alpha$ expansion of the external field approximation
\cite{Yen90,Eid01})
\[
\begin{array}{@{}l}\displaystyle
\delta E_{1\mbox{-}loop}^{(4)} = \alpha^4
    \biggl[
       4\pi\left(\frac{139}{128}-\frac{1}{2}\ln{2}\right)
       +\frac{5\pi}{48}
    \biggr]
    \Bigl\langle
       \delta(\mathbf{r}_1)+\delta(\mathbf{r}_2)
    \Bigr\rangle,
\end{array} \]
and two-loop QED corrections,
\[
\delta E_{2\mbox{-}loop}^{(4)} = \frac{\alpha^4}{\pi}
\left[-\frac{4358}{1296}-\frac{10\pi^2}{27}
+\frac{3\pi^2}{2}\ln{2}-\frac{9}{4}\zeta(3)\right]
\Bigl\langle
   \delta(\mathbf{r}_1)\!+\!\delta(\mathbf{r}_2)
\Bigr\rangle.
\]

The last contribution is the relativistic corrections of order $\alpha^4$,
$\delta E^{(4)}_{rc}$, for the bound electron (the $m\alpha^6$ order term
in the expansion of the Dirac energy for the two--center problem).
\[
\delta E^{(4)}_{rc} =
   \left\langle H_B Q (E_0-H_0)^{-1} Q H_B \right\rangle
  +\left\langle H^{(4)} \right\rangle
\]
where $H_B$ is the Breit-Pauli Hamiltonian for the bound electron of the
two center problem and $Q$ is a projector orthogonal to the initial
$2p\sigma_u$ electronic state, and
\[
\begin{array}{@{}l}
\displaystyle
H^{(4)} = \alpha^4\frac{p_e^6}{16m^5}
     +\alpha^4\sum_{i=1,2}\biggl(\frac{1}{8m^3}
         \left[\mathbf{p}_e,\frac{1}{r_i}\right]
\\[4mm]\displaystyle\hspace{15mm}
     \!-\!\frac{3\pi}{16m^4}
        \left\{
           p^2_e,\left[\mathbf{p}_e,
               \left[\mathbf{p}_e,\frac{1}{r_i}\right]\right]
        \right\}
     \!+\!\frac{5}{128m^4}\left[p^2_e,\left[p^2_e,\frac{1}{r_i}\right]
  \right]\biggr).
\end{array}
\]
In fact, the $\alpha^4$ relativistic correction turns out to be
negligibly small and can be omitted.

\begin{table}
\begin{center}
\begin{tabular}{lr@{}l}
\hline\hline
$\Delta E_{nr}$          & $-$1.&085$\>045\>252(1)\times10^{-9}$ \\
$\Delta E_{\alpha^2}$    &    0.&003$\>298\>5(4)\times10^{-9}$ \\
$\Delta E_{\alpha^3}$    & $-$0.&000$\>470\>02(1)\times10^{-9}$ \\
$\Delta E_{\alpha^4}$    & $-$0.&000$\>003\>29\times10^{-9}$ \\
\hline
$E_{B}$         &    1.&082$\>219\>8(4)\times10^{-9}$ \\
\hline\hline
\end{tabular}
\end{center}
\caption{Relativistic and QED corrections to the $2p\sigma_u(v=1)$ state of
the hydrogen molecular ion $\mbox{H}_2^+$.}\label{Table3}
\end{table}

The summary of the relativistic and QED contributions up to and including
term of order $\alpha^4$ is presented in Table \ref{Table3}. The
uncertainty in the final value is determined by the uncertainty in
calculating the leading relativistic correction for the bound electron.
The other corrections have been obtained with much better accuracy.

\section{Spin effects}

In the preceding evaluations,  the spin  effects were ignored. For the
$2p\sigma_u(v=1)$ state, the spin-spin interaction has a form:
\[
H^{\rm HFS}_{{\rm H}_2^+} = \alpha^2\frac{8\pi}{3}
      \mu_e\mathbf{s}_e
      \Bigl[\mu_p\mathbf{s}_{p_1}\delta(\mathbf{r}_1)
           +\mu_p\mathbf{s}_{p_2}\delta(\mathbf{r}_2)\Bigr],
\]
where $\mu_e=(1+a_e)/m_e$ and $\mu_p=(1+a_p) /M_p$ are the magnetic moments
of an electron and a proton, respectively. For the lowest hyperfine state,
$S_{\rm tot}=1/2$, the spin-dependent correction to the binding energy is
\[
\delta E^{\rm HFS}_{{\rm H}_2^+} = \alpha^2\frac{8\pi\mu_e\mu_p}{3}
      \bigl\langle\mathbf{S}_n\mathbf{s}_e\bigr\rangle
      \Bigl[
          \left\langle
             \delta(\mathbf{r}_1)+\delta(\mathbf{r}_2)
          \right\rangle_{{\rm H}_2^+}
      \Bigr],
\]
where $\mathbf{S}_n=\mathbf{s}_{p_1}+\mathbf{s}_{p_2}$ is a total spin of
protons, and $\bigl\langle\mathbf{S}_n\mathbf{s}_e\bigr\rangle=-1$. On the
other hand, the $p+\mbox{H}$ asymptotic states should be antisymmetric
with respect to exchange of two protons as well. For the $2p\sigma_u(v=1)$
this can be realized only when proton spins are parallel ($S_n=1$).  In
this case the threshold energy should be
\[
E^{\rm HFS}_{\rm H}(S_{\rm tot}=1/2) = -\alpha^2\frac{8\pi\mu_e\mu_p}{3}
      \Bigl[
          \left\langle\delta(\mathbf{r})\right\rangle_{{\rm H}}
      \Bigr].
\]
Then the change of the binding energy due to the spin-spin interaction is
\[
\Delta E^{\rm HFS}_{S_{\rm tot}=1/2} =
    \left(
       \delta E^{\rm HFS}_{{\rm H}_2^+}
       -E_{\rm H}^{\rm HFS}
    \right)_{S_{\rm tot}=1/2}
   = -8.223\times10^{-14} \mbox{ a.u.}
\]

\section{Conclusions}

The relativistic and QED corrections to the $2p\sigma_u(v=1)$
vibrational state of the $\mbox{H}_2^+$ molecular ion have been evaluated.
Calculations include up to $\alpha^4$ order terms.

The main conclusion of this work is that, despite the smallness of its
binding energy, the existence of this state is not questioned by the
impact of the relativistic and radiative effects. By including  all these
corrections one gets a binding energy $E_B=1.082219 8(4) \times 10^{-9}$
a.u., or $E_B=2.944\>870(1)\times 10^{-8}$ eV, what represents a relative
modification $\Delta E_B/E_B= 2.61 \times 10^{-3}$ of the nonrelativistic
value.

While the nonrelativistic binding energy is now known to a relative
precision of about $10^{-9}$, the final value for the physical binding
energy has an uncertainty by two orders of magnitude larger.
Generally, for the low vibrational "gerade" states of the hydrogen isotope
molecular ions, the convergence with respect to $\alpha$ is
better. The most conceivable explanation is that, for the weakly bound
states, the cancelation of the different correction terms in the molecular
ion and the atom has the strongest effect and this  slows down the
convergence of the expansion. It  manifests already in the leading
order corrections of the Breit-Pauli Hamiltonian and was observed
before in \cite{How90} for the $1s\sigma_g$ series of states.

The contribution of the spin-spin interaction is rather small: its
magnitude is by three orders smaller than the relativistic correction for
the bound electron.

Finally, we would like to mention that the existence of this weakly bound
$\mbox{H}_{2}^{+}$ molecular ion state is of fundamental importance. It
manifests itself in a huge $p$--H scattering length $a\approx750$~a.u.
\cite{Car02}, which determines the low energy scattering of proton by
atomic hydrogen. The $\mbox{H}_{2}^{+}$ formation rate is substantially
influenced by the $p$--H resonant cross section. This can help to explain
the abnormal abundance of H$_2$ molecules  in the interstelar space
\cite{Smukstaras,Smukstaras1}. Experimental confirmation of the considered
state would be very appreciated, despite being difficult to realize
\cite{Car03}.

\section{Acknowledgements}

One of the authors (V.K.) wants to acknowledge the support from the
Laboratoire de Physique Subatomique et de Cosmologie and express deep
thanks to J.-M.~Richard for generous hospitality and useful discussions.

\end{document}